\documentclass[aip, twocolumn, reprint, cha]{revtex4-1}
\usepackage[utf8]{inputenc}
\usepackage[T1]{fontenc}
\usepackage{dcolumn} 
\usepackage{graphicx}
\usepackage{latexsym}
\usepackage{pstricks, pst-node, pst-text, pst-3d}
\usepackage{amssymb, amsmath, amscd}
\usepackage{mathrsfs, mathtools}
\usepackage{verbatim}
\usepackage{hyperref}
\usepackage{placeins}
\usepackage{multirow}
\usepackage{pifont}
\usepackage{lipsum}
\usepackage{array}
\usepackage{color}
\usepackage{bm}
\hypersetup{
	colorlinks = true, %colors links instead of ugly boxes
	urlcolor   = magenta, %color for external hyperlinks
	linkcolor  = red, %color of internal links
	citecolor  = blue %color of citations
}
\definecolor{azul}{rgb}{0,0.1,0.9}
\definecolor{texto}{rgb}{0,0,0}
\definecolor{cerceta}{rgb}{0,0.5,0.5}
\definecolor{vermelho}{rgb}{0.95,0.,0.05}

\begin{document}

\title{Impact of periodic vaccination in SEIRS seasonal model}

\author{Enrique C. Gabrick} 
\email{enriquec@pik-potsdam.de;\\ecgabrick@gmail.com}
\affiliation{Potsdam Institute for Climate Impact Research, Telegrafenberg A31, 14473 Potsdam, Germany.}
\affiliation{Department of Physics, Humboldt University Berlin, Newtonstraße 15, 12489 Berlin, Germany.}
\affiliation{Graduate Program in Science, State University of Ponta Grossa, 84030-900, Ponta Grossa, PR, Brazil.}
\author{Eduardo L. Brugnago}
\affiliation{Institute of Physics, University of S\~ao Paulo, 05508-090, S\~ao Paulo, SP, Brazil.}
\author{Silvio L.T. de Souza}
\affiliation{Federal University of S\~ao Jo\~ao del-Rei, Campus Centro-Oeste, 35501-296, Divin\'opolis, MG, Brazil.}
\author{Kelly C. Iarosz} 
\affiliation{Graduate Program in Science, State University of Ponta Grossa, 84030-900, Ponta Grossa, PR, Brazil.}
\affiliation{University Center UNIFATEB, 84266-010, Tel\^emaco Borba, PR, Brazil.}
\author{Jos\'e D. Szezech Jr.}
\affiliation{Graduate Program in Science, State University of Ponta Grossa, 84030-900, Ponta Grossa, PR, Brazil.}
\affiliation{Department of Mathematics and Statistics, State University of Ponta Grossa, 84030-900, Ponta Grossa, PR, Brazil.}
\author{Ricardo L. Viana}
\affiliation{Institute of Physics, University of S\~ao Paulo, 05508-090, S\~ao Paulo, SP, Brazil.}
\affiliation{Department of Physics, Federal University of Paran\'a, 81531-980, Curitiba, PR, Brazil.}
\author{Iber\^e L. Caldas}
\affiliation{Institute of Physics, University of S\~ao Paulo, 05508-090, S\~ao Paulo, SP, Brazil.}
\author{Antonio M. Batista}
\affiliation{Graduate Program in Science, State University of Ponta Grossa, 84030-900, Ponta Grossa, PR, Brazil.}
\affiliation{Institute of Physics, University of S\~ao Paulo, 05508-090, S\~ao Paulo, SP, Brazil.}
\affiliation{Department of Mathematics and Statistics, State University of Ponta Grossa, 84030-900, Ponta Grossa, PR, Brazil.}
\author{J\"urgen Kurths}
\affiliation{Potsdam Institute for Climate Impact Research, Telegrafenberg A31, 14473 Potsdam, Germany.}
\affiliation{Department of Physics, Humboldt University Berlin, Newtonstraße 15, 12489 Berlin, Germany.}

\begin{abstract}
{We study three different strategies of vaccination in a SEIRS (Susceptible--Exposed--Infected--Recovered--Susceptible) 
	seasonal forced model, 
	which are: 
	($i$) continuous vaccination; 
	($ii$) periodic short time localized vaccination and 
	($iii$) periodic pulsed width campaign. 
	Considering the first strategy,} 
we obtain an expression for the basic reproduction number and 
infer a minimum vaccination rate necessary to ensure the stability of 
the disease-free equilibrium (DFE) solution. 
{In the second strategy, 
	the short duration pulses are added to a constant baseline vaccination rate. 
	The pulse is applied according to the seasonal forcing phases. 
	The best outcome is obtained by locating the intensive immunization at inflection of the transmissivity curve. 
	There, 
	a vaccination rate of $44.4\%$ of susceptible individuals is enough to ensure DFE.} 
{For the third vaccination proposal, 
	additionally to the amplitude, 
	the pulses have a prolonged time width. 
	We obtain a non-linear relationship between vaccination rates and 
	the duration of the campaign. 
	Our simulations show that the baseline rates, 
	as well as the pulse duration, 
	can substantially improve the vaccination campaign effectiveness. 
	These findings are in agreement with our analytical expression.} 
We show a relationship between the vaccination parameters and 
the accumulated number of infected individuals, 
over the years and 
show the relevance of the immunisation campaign annual reaching for controlling the infection spreading. 
{Regarding the dynamical behaviour of the model, 
	our simulations shows that chaotic and  
	periodic solutions, 
	as well as bi-stable regions, 
	depend on the vaccination parameters range.}
\end{abstract}

\maketitle

\begin{quotation}
The spread of infectious diseases is a challenge to world public health. 
Many efforts are dedicated to mitigating the impacts of the spreading of diseases. 
In this context, 
mathematical model is a powerful tool to simulate, 
forecast and 
study efficient control measures for human and 
wildlife diseases. 
Although there are many types of disease, 
some of them have common characteristics, 
for instance to repeat the peak of infectious in certain times. 
These diseases are called seasonal, 
e.g., 
measles, 
mumps and 
smallpox. 
The reasons for the seasonality can be varied, 
such as climate and social behaviours. 
Due to the seasonal characteristic of these diseases, 
they recur in the populations and 
control measures are needed to be implemented in order to eradicated them. 
One the most successful control measure is a vaccination campaign, 
that can be continuous or periodic. 
{Continuous vaccination is a campaign in which 
	a constant quantity of vaccines is daily available to the population throughout the year, 
	while the periodic strategy considers an intense immunization campaign, 
	during a fraction of the year.}  
In this work, 
we study the impacts of vaccination in a SEIRS model with seasonal forcing. 
We consider newborns vaccine and susceptible vaccination, 
administrated in three different ways: 
($i$) continuous vaccination; 
($ii$) {periodic short time localized vaccination and 
($iii$) periodic pulsed width campaign.} 
For the periodic strategies, 
we consider a baseline rate. 
{In the constant vaccine context,} 
we obtain an analytical expression for the vaccination rate in which the (DFE) is guaranteed.  
For the parameters considered, 
the annually vaccinated value is in agreement with numerical simulations for all the vaccine strategies. 
Furthermore, 
considering parameters for which bi-stability exist in the case without vaccine, 
our results show that it persists depending on the vaccination rate. 
\end{quotation}

%%%%%%%%%%%%%%%%%%%%%%%%%%%%%%%%%%%%%%%%%%
%%%%%%%%%%%%%%%%%%%%%%%%%%%%%%%%%%%%%%%%%%
\section{Introduction}
Infectious diseases spreading is a highly important problem in public health~\cite{Xia2023}. 
The spread of infectious diseases affects the humanity throughout whole history~\cite{Glatter2021}, 
e.g., 
Black Death during the fourteenth century~\cite{Glatter2021}, 
Spanish flu in 1918~\cite{Tumpey2005}, 
COVID-19~\cite{Machein2020}, 
Dengue Fever~\cite{Aguiar2011}, 
HIV~\cite{Dalal2008}, etc.~\cite{Meyers2005, Mao2010, Scarpino2019}. 
Some of these illness present seasonal patterns of incidence~\cite{Altizer2006}, 
namely seasonal infections diseases~\cite{Buonomo2018}. 
The mechanisms of seasonality are varied, 
such as weather~\cite{Hoshen2004}, 
school holidays~\cite{Finkenstadt2000} and 
others~\cite{Grassly2006}. 
Examples of seasonal infectious diseases are mumps~\cite{Azimaqin2022}, 
measles~\cite{Xia2023} and 
dengue fever~\cite{Aguiar2009}. 
An introduction to corresponding models is found in Refs.~\cite{Altizer2006, Buonomo2018}. 

Mathematical models are a powerful tool to understand, 
forecast and 
study diseases spread control measures~\cite{Mugnaine2022}. 
In epidemiology, 
the standard models are compartmental~\cite{Batista2021}, 
where the host population is compartmentalized according to the stages of the infection evolution and 
the possible states considered. 
The individuals {in the host population} are taken from susceptible to possible intermediate stages, 
as infectious and 
recovered.  
Depending on the disease to be modelled, 
when {an infection} cycle is completed, 
the individuals can {acquire permanent} immunity~\cite{Grenfell1995}, 
{or become susceptible again.}
These models are easily adapted to study different diseases, 
as seasonal infectious ones~\cite{Keeling2008}, 
which can be modelled by the inclusion of a nonlinear time-dependent term in the transmissivity, 
e.g., 
a sinusoidal forcing~\cite{Bilal2016} or a square wave function~\cite{Tanaka2013}. 
Due to this non-linearity, 
the resulting dynamics can become very {intricate}, 
even exhibiting chaos~\cite{Olsen1990} or bi-stability~\cite{Gabrick2023}. 

From a modelling perspective, 
seasonal models have been used since 1928~\cite{Buonomo2018}. 
These models reproduce with great accuracy the dynamics observed in diseases, 
such as measles~\cite{Olsen1990}, 
chickenpox, 
mumps~\cite{London1973} and 
others~\cite{Altizer2006}. 
In a formulation of the SIRS (Susceptible--Infected--Recovered--Susceptible) 
model with seasonal contact rate, 
Greenhalgh and Moneim~\cite{Greenhalgh2003} 
showed the existence of an unique DFE solution, 
which is globally asymptotically stable when the basic reproduction number is less than one ($\mathcal{R}_0<1$), 
{where} $\mathcal{R}_0$ is a measure of the reproductive potential for a given disease. 
In a population, 
where everyone is initially susceptible, 
the {infection} can {remain} only if $\mathcal{R}_0>1$~\cite{Keeling2008}. 
Furthermore, 
Greenhalgh and Moneim 
considered four childhood infectious diseases 
(measles, chickenpox, mumps, and rubella) and 
showing non-trivial solutions (chaotic dynamics). 
These {results} are found by analysing the bifurcation diagram, 
where some ranges exhibit chaotic attractors. 
Also in a SIR (Susceptible--Infected--Recovered) framework, 
de Carvalho and Rodrigues~\cite{Carvalho2022} 
{implemented} a multi-parameter periodically forced term leading to strange attractor solutions. 
In their formulation, 
the DFE is not preserved when $\mathcal{R}_0 < 1$. 
Metcalf et al.~\cite{Metcalf2009} studied the seasonal variation of six childhood infections 
(measles, pertussis, mumps, diphtheria, varicella and scarlet fever) 
from data of Copenhagen in the pre-vaccination era. 
Their results showed that the transmission disease decreases for some infections at school holidays. 

Previous works reported the capacity of compartmental models to reproduce {chaotic} dynamics~\cite{Keeling2001} and 
the accuracy in simulating real data~\cite{Machein2020, Cooper2020}. 
However, 
one of the most important advantages in using compartmental models is the facility to {implement} important characteristics to simulate control measures, 
such as social distancing~\cite{Ansari2021, Mello2020}, 
restrictive measures~\cite{Mugnaine2022, Brugnago2020, Souza2021}, 
quarantine~\cite{Balsa2021, Zou2022}, 
vaccination~\cite{Etxeberria-Etxaniz2020, DelaSen2021}, 
etc. 
Despite there are some forms of control, 
one {of} the most effective is vaccination~\cite{Voysey2021, Thompson2022, Zou2022}. 

Vaccination campaign in infectious seasonal diseases shows a 
significant infected number decrease~\cite{Grassly2006, Molinari2007, Keeling2002}. 
Gao et al.~\cite{Gao2011} studied the vaccination of newborns combined with pulsed vaccine in a SIRS seasonal forced model, 
in a modelling approach. 
One of the results obtained in this research is the DFE when $\mathcal{R}_0 < 1$. 
The rotavirus vaccination was studied by Atchison et al.~\cite{Atchison2010} 
using a modified SIR model to fit the data from England and Wales. 
From these simulations, 
they reported that vaccination reduces rotavirus diseases transmission by $61\%$. 
Considering a pulsed vaccination strategy in a SIR model, 
Shulgin et al.~\cite{Shulgin1998} showed the possibility of disease eradication with relatively low {vaccination rates}. 
To get the disease eradication, 
they explored some conditions, 
as vaccine proportion and 
periodicity. 
The effects of two vaccination doses in a SEIR (Susceptible-Exposed-Infected-Recovered) 
epidemic model was studied by Gabrick et al.~\cite{Gabrick2022}. 
In this work, 
they considered three vaccination strategies: 
unlimited doses applied continuously in the susceptible population; 
limited doses supply applied periodically in the susceptible population; 
and limited doses applied in a periodic strategy in all host population. 
Their results showed that the vaccine campaign is more efficient when applied only in susceptible individuals, 
i.e., 
the population is previously tested. 
Considering a Kot-type function as seasonality, 
Duarte~\cite{Duarte2021} analysed the control of infectious disease with vaccination strategies including a perturbation term. 
Seasonal contact and 
optimal vaccination strategy were studied by Wang~\cite{Wang2019} in a SEIR  model. 
The persistence or extinction of a seasonal disease with reinfection possibility was discussed by Bai and Zhou~\cite{Bai2012}. 
They explored the conditions for the extinction of the diseases in the situations where $\mathcal{R}_0<1$ and,  
{also} $\mathcal{R}_0>1$. 
Their results showed that $\mathcal{R}_0 = 1$ is the threshold for the disease extinction. 
However, 
from simulations, 
a policy only based on $\mathcal{R}_0$ can overestimate the infections risks and 
the infected number due the presence of seasonality. 
Periodic strategy vaccination in a SEIRS model was discussed by Moneim and 
Greenhalgh~\cite{Moneim2005}. 
The vaccine periodicity was described as integer multiples of the contact rate period. 
They reported that a key parameter to understand the vaccine influence is the $\mathcal{R}_0$. 
Also, 
they proved that $\mathcal{R}_0<1$ is associated with the DFE and 
is globally asymptotically stable.
Other works that address the vaccination of seasonal diseases can be found in 
Refs.~\cite{Ho2019, Metcalf2012, Nokes1995, Agur1993} and 
in the references therein. 

In this work, 
we study the effects of vaccination in a SEIRS seasonal forced model \cite{Bai2012, Gabrick2023}. 
We substantially extended the works of Moneim and Greenhalgh \cite{Moneim2005} and 
Gabrick et al. \cite{Gabrick2023}. 
The first one explored the effects of periodic vaccination and 
conjectured one $\mathcal{R}_0$ as a function of vaccine parameters. 
However, 
the authors did not explore the effects of constant vaccine and 
periodic pulses with different width. 
In the second one, 
the authors obtained a parameter range in which the numerical solutions exhibit bi-stability and 
tipping point phenomena can be explored. 
However, 
{they} did not taken into account the vaccine influences in the dynamical behaviour. 
In this way, 
our work extend the Ref.~\cite{Moneim2005} {by implementing} 
{three different strategies of vaccination in susceptible and 
	newborns individuals, 
	namely 
	($i$) constant, 
	($ii$) pulsed and 
	($iii$) pulsed width vaccination strategies.}
{We choose these three strategies because:  
	($i$) it can be used to model mass vaccination programs~\cite{Grabenstein2006}; 
	($ii$) it can be considered to model a scenario in which a certain amount of vaccination (baseline) 
	is available during all the time, 
	but periodically an intense campaign are employed~\cite{Alkhamis2022}; 
	and strategy ($iii$) modified ($ii$) by extending the duration 
	of the intensive immunization campaign to days, 
	weeks or months, 
	i.e., 
	it saves vaccination efforts.} 
{Then,} 
considering the parameters found in Ref.~\cite{Gabrick2023} for bi-stability, 
we explore, 
as novelty, 
the dynamical aspects of the {SEIRS seasonal} in the presence of vaccine, 
{according to the three strategies mentioned.} 
Our simulations show that for the 
($i$) case the vaccine in susceptible is more significant and 
rates $\geq 44\%$ are able to extinct the infection in the host population. 
In the ($ii$) scenario we found a linear relationship between the baseline and 
pulsed immunization rate {to result in the infection eradication}. 
Also, 
acting the pulses at the inflection point of the seasonality function, 
the campaign is slightly more effective. 
To the strategy ($iii$), 
we discover a non-linear relationship between the immunization {campaign} parameters that leads to the illness extinction. 
   
The current work is organised as follows: 
in Sec.~\ref{sec_modelo} we present the model and 
its equilibrium solutions, 
from which we obtain the $\mathcal{R}_0$ as well as a minimum rate to eradicate the disease. 
In Sec.~\ref{sec_vacinaConstante} we discuss the effects of constant vaccination in the model. 
The pulsed vaccination strategy is discussed in Sec.~\ref{sec_vacinaPulsada} and  
the implementation of pulsed width in the vaccine {campaign} is present in Sec.~\ref{sec_vacinaPulsoLargo}. 
Finally, 
our conclusions are drawn in Sec.~\ref{sec_conclusion}. 

%%%%%%%%%%%%%%%%%%%%%%%%%%%%%%%%%%%%%%%%%%%%%%%
%%%%%%%%%%%%%%%%%%%%%%%%%%%%%%%%%%%%%%%%%%%%%%%
\section{Model} \label{sec_modelo}
\begin{figure}[!h]
	\centering
	\includegraphics[width=1.\columnwidth]{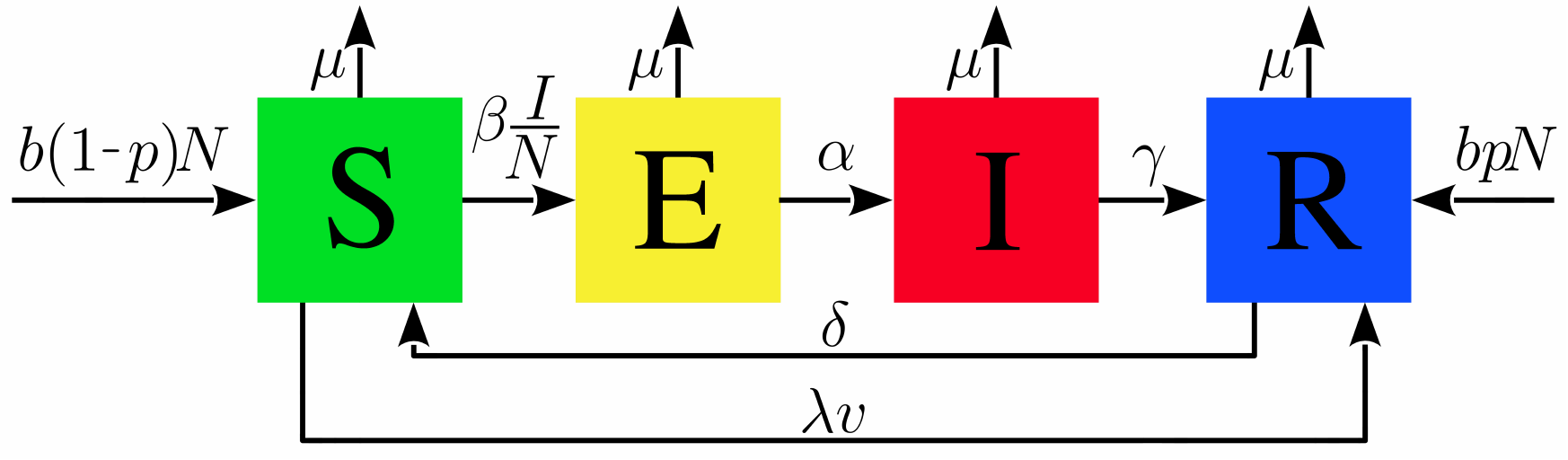}
	\caption{Schematic representation of SEIRS model with vaccine inclusion. 
	The compartments (coloured boxes), 
	indexed by variables of the system~(\ref{eq1}), 
	contains fractions of the population at each stage of infection spread. 
	Arrows indicate the flow between them, 
	increase in newborns and 
	loss of population due to death;  
	accompany the respective transition rates.} 
	\label{fig1}
\end{figure}

SEIRS is a compartmental epidemiological model that 
describes the spread of a given infectious disease in a 
homogeneously mixed host population,  
in which individuals are computed into one of four compartments,
namely susceptible in $S$, 
exposed and not yet contagious in $E$, 
infectious in $I$ and recovered in $R$~\cite{Keeling2008}. 
The population size ({$N=S+E+I+R$}) 
is time-dependent when  the natural death rate ($\mu$) 
is not equal to the birth rate ($b$). 
The transition flow between the compartments is schematically represented in Fig.~\ref{fig1}. 
Susceptible individuals become infected through interaction with contagious agents at a rate $\beta{I/N}$, 
where $\beta$ is the transmitting infection rate per interaction. 
Once a portion of $S$ is infected, 
it evolves to $E$, 
remaining in this compartment by an average time given by $1/\alpha$ (latent period). 
The parameter $\alpha$ is the rate at which exposed individuals become infectious, 
migrating to compartment $I$. 
Individuals remain in $I$ by an average time $1/\gamma$ (infectious interval),  
after that they occupy the $R$ compartment, 
the parameter $\gamma$ is called recovery rate. 
In this model there is no permanent immunity, 
then, 
after an average interval $1/\delta$ the individuals in $R$ return to $S$~\cite{Batista2021}. 
In which 
$\delta$ is the immunity loss rate. 
Furthermore, 
we consider vaccination of newborns and 
susceptible~\cite{Moneim2005}. 
A fraction {$p\in[0,1]$} of newborns are directly immunized and 
a portion of $S$ is vaccinated at a rate $v$ with effectiveness $\lambda\in[0,1]$. 
The vaccine effect is to give immunity to individuals, 
transferring, 
in this model, 
newborn and 
susceptible ones to the recovered class. 
{Given the nature of the model, 
	all parameters are non-negative real numbers.} 

The SEIRS model with vaccination, 
as described above, 
is given by the following system of four coupled ordinary differential equations \cite{Moneim2005}: 
\begin{equation} 
	\begin{split}
		\frac{dS}{dt} & = b(1-p)N + \delta R - \left(\beta\frac{I}{N} + \lambda v + \mu\right) S, \\
		\frac{dE}{dt} & = \beta \frac{SI}{N} - (\alpha + \mu)E, \\
		\frac{dI}{dt} & = \alpha E - (\gamma + \mu)I, \\
		\frac{dR}{dt} & = bpN + \lambda vS + \gamma I - (\delta + \mu)R.
	\end{split}
	\label{eq1}
\end{equation}
{Other effects also can be considered in this model and 
	incorporated into Eqs.~(\ref{eq1}), 
	such as reaction-diffusion~\cite{Song2019} and delay~\cite{Cooke1996}, 
	but this is beyond the present work and 
	will be studied next.} 
Seasonality is included in the model by replacing $\beta$ by a periodic function: 
\begin{equation}
	\beta(t) = \beta_0 \left[1 + \beta_1 {\rm cos} (\omega t)\right],
	\label{eq_betaFuncao}
\end{equation} 
where $\beta_0$ is the average contagion rate, 
the seasonality degree is $\beta_1\in [0,1]$, and 
$\omega$ is its frequency~\cite{Olsen1990}. 

From Eqs.~(\ref{eq1}), 
the sum of variables followed by a simple manipulation, 
provides us the growth rate of the host population $dN/dt = (b - \mu)N$. 
Considering this result, 
the Eqs.~(\ref{eq1}) can be rewritten in a normalised form, 
without loss of generality, 
by taking~\cite{Bjornstad2018} 
\begin{equation*}
	S = Ns,~E = Ne,~I = Ni,~\text{and}~R = Nr,
\end{equation*}
which gives the respective transformations for the time derivatives of the variables: 
\begin{equation}
	\begin{split}
		\frac{ds}{dt} &= \frac{1}{N}\left(\frac{dS}{dt} - (b - \mu)S\right),\\[1.5pt]
		\frac{de}{dt} &= \frac{1}{N}\left(\frac{dE}{dt} - (b - \mu)E\right),\\[1.5pt]
		\frac{di}{dt} &= \frac{1}{N}\left(\frac{dI}{dt} - (b - \mu)I\right),\\[1.5pt]
		\frac{dr}{dt} &= \frac{1}{N}\left(\frac{dR}{dt} - (b - \mu)R\right). 
	\end{split}
	\label{eq_transformaVariaveis}
\end{equation}
This change of variables leads to equations of the same form that would be obtained by assuming $b = \mu$. 
We get 
\begin{equation} 
	\begin{split}
		\frac{ds}{dt} &= b(1-p) + \delta r - \left[ \beta(t) i + \lambda v + b \right] s, \\
		\frac{de}{dt} &= \beta(t) is - (\alpha + b)e, \\
		\frac{di}{dt} &= \alpha e - (\gamma + b)i, \\
		\frac{dr}{dt} &= bp + \lambda vs + \gamma i - (\delta + b)r, 
	\end{split}
	\label{eq3}
\end{equation}
with the constrain $s + e + i + r = 1$.
Then, 
$r$ can be determined in terms of the other three variables, 
such that we replace it in the first of Eqs.~(\ref{eq3}), 
reducing the model to the following three-equation system: 
\begin{align} 
	\frac{ds}{dt} &= b(1-p) + \delta  - \left[ \beta(t) i + \lambda v + \delta + b  \right] s - \delta (e + i), \nonumber \\
	\frac{de}{dt} &= \beta(t) is - (\alpha + b)e, \label{eq_sistemaNormalizado} \\
	\frac{di}{dt} &= \alpha e - (\gamma + b)i. \nonumber
\end{align}

%%%%%%%%%%%%%%%%%%%%%%%%%%%%%%%%%%%%%
\subsection{{Disease-free equilibrium}}
In general, 
Eqs.~(\ref{eq_sistemaNormalizado}) are solved numerically. 
However, 
for the particular case when $\beta = \beta_0$ is constant, 
there are two fixed point solutions, 
which are: 
DFE and endemic solution~\cite{Guillen2017}. 
The first one is characterised by the disappearance of the disease in the host population ($i^* = 0$); 
in the other, 
the infection remains ($i^* > 0$). 
Firstly, 
we investigate the DFE solution given by 
\begin{equation}
	(s^*,e^*,i^*) = \left( \frac{b(1-p) + \delta}{\lambda v + \delta + b }, 0 ,0 \right).  
	\label{eq_DFE}
\end{equation}
This fixed point is stable if all eigenvalues of the Jacobian matrix ($\mathbb{J}$) of the system, 
computed in DFE, 
have negative real part~\cite{TamasTel}. 
Given the matrix 
\begin{equation}
	\mathbb{J}_{\rm DFE} = -\left[
	\begin{array}{ccc}
		(\lambda v + \delta + b)  & \delta  & (\beta_0 s^* + \delta)  \\[4pt]
		 0 & (\alpha + b)  & -\beta_0 s^* \\[4pt]
		 0 & -\alpha  & (\gamma + b)
	\end{array}
	\right], 
\end{equation} 
its eigenvalues are 
\begin{align}
	\xi_1 &= - (\lambda v + \delta + b), \\[4pt]
	\xi_2 &= \frac{-(\alpha + \gamma + 2b) - \sqrt{(\alpha - \gamma)^2 + 4 \alpha \beta_0 s^*}}{2}, \\[4pt]
	\xi_3 &= \frac{-(\alpha + \gamma + 2b) + \sqrt{(\alpha - \gamma)^2 + 4 \alpha \beta_0 s^*}}{2}. 
\end{align}
All of them are real numbers, 
once the constants are positive and 
$\xi_{1,2}$ are always negative, 
then the DFE is stable when $\xi_{3}<0$. 
From this inequality, 
we obtain the relation for the stability of this fixed point: 
\begin{equation}
	\mathcal{R}_0 = \frac{\alpha \beta_0 s^*}{(\alpha + b)(\gamma + b)} < 1. 
	\label{eq_r0}
\end{equation}
Therefore, 
if $\mathcal{R}_0 < 1$ the DFE point is stable, 
i.e. the disease will die out. 
On the other hand,  
if $\mathcal{R}_0 > 1$ the disease will be succeed in infect the population~\cite{AndersonMay}. 
Thus, 
we establish a stability condition for the DFE relative to the basic reproduction number.
The quantity $\mathcal{R}_0$ informs about the evolution of the spread. 
Note that, 
if $p = \lambda v = 0$, 
then $s^* = 1$ and 
we recover the $\mathcal{R}_0$ for the model without vaccine~\cite{Gabrick2023}.

Note that Eq.~(\ref{eq_r0}) is for the autonomous case,  
{with $\beta$ and} 
{$v$ is constant}. 
However, 
for periodic diseases {without latent period}~\cite{Allen2010} the $\mathcal{R}_0$ 
expression can be extend replacing $\beta_0$ by $\langle{\beta (t)}\rangle$: 
\begin{equation}
	\langle{\beta (t)}\rangle = \frac{1}{T_{\rm f} - T_0} \int_{T_0}^{T_{\rm f}} \beta(t) dt, 
	\label{beta_medio}
\end{equation}
{where $T_{\rm f} - T_0$ corresponds to the seasonality period.} 
{In the case studied in this work, 
	we can analyze both the average $\langle \mathcal{R} \rangle$ of the time-dependent reproduction number and 
	its maximum $\mathcal{R}_{+}$ within a seasonal cycle~\cite{Moneim2005}, 
	being $\langle \mathcal{R} \rangle = \mathcal{R}_0$ and 
	$\mathcal{R}_{+} = (1 + \beta_1)\mathcal{R}_0$.} 
{imposing the stable DFE condition on these quantities, 
	we obtain two quotas:}
\begin{align}
	 \frac{\alpha [b(1-p) + \delta]}{\lambda(\alpha + b)(\gamma + b)}\beta_0 - \frac{(\delta + b)}{\lambda} &< v_0,\\ 
	 \frac{\alpha [b(1-p) + \delta]}{\lambda(\alpha + b)(\gamma + b)}\beta_0(1 + \beta_1) - \frac{(\delta + b)}{\lambda} &< v_+. 
\end{align}

{The disease extinction is guaranteed when the relationship $v_+ \leq v$ is verified.} 
{However, 
	this upper limit for the vaccination rate can be greater than enough, 
	with $v_0 \leq v$ being sufficient to reach the DFE in our simulations. 
	Note that this threshold is the same in the autonomous case.} 
As a numerical example, 
considering
$\lambda = 1$,
$p = 0.25$
$b=0.02$, 
$\beta_0 = 270$, 
$\delta=0.25$, 
$\alpha=100$ and
$\gamma=100$, 
we obtain a {quota} $v_0 = 0.445$. 
Therefore, 
for a constant vaccination campaign the minimum rate is $44.5\%$ of vaccinated susceptible. 
{A value very close to those numeriacally obtained in Sec.~\ref{sec_vacinaConstante} and 
	annual vaccination rates in Secs.~\ref{sec_vacinaPulsada} and 
	\ref{sec_vacinaPulsoLargo}.}

%%%%%%%%%%%%%%
\subsection{{Endemic equilibrium}}
{Considering the autonomous case,} 
the endemic equilibrium point {${\rm EE}\left(\widetilde{s}, \widetilde{e},  \widetilde{i} \right)$ } is given in terms of $\mathcal{R}_0$ and 
$s^*$, as 
\begin{align}
	%\widetilde{s} &= \frac{s^*}{\mathcal{R}_0}, \\[1.5pt] 
	\widetilde{s} &= \frac{(\alpha + b)(\gamma + b)}{\alpha\beta_0} = \frac{s^*}{\mathcal{R}_0}, \\[1.5pt] 
	\widetilde{e} &= \frac{(\gamma + b) \left[ b(1 - p) + \delta \right]}{\alpha \beta_0 s^* + \delta(\alpha + \gamma + b)\mathcal{R}_0}
	\left(\mathcal{R}_0 - 1 \right), \label{eq_eEndemico}\\[1.5pt] 
	\widetilde{i} &= \frac{\alpha \left[ b(1 - p) + \delta \right]}{\alpha \beta_0 s^* + \delta(\alpha + \gamma + b)\mathcal{R}_0}
	\left(\mathcal{R}_0 - 1 \right). \label{eq_iEndemico}
\end{align}
In this solution the disease is permanent in the host population. 
It is worth to note that for the inverse relation {$\widetilde{s}$ proportional to $\mathcal{R}_0^{-1}$}, 
the scenario without vaccination verifies the result~\cite{Gabrick2023} $\widetilde{s} = \mathcal{R}_0^{-1}$. 
This fixed point exists only if $\mathcal{R}_0 > 1$, 
since $\mathcal{R}_0 < 1$ implies $\widetilde{e}, 
\widetilde{i} < 0$ in Eqs.~(\ref{eq_eEndemico}) and~(\ref{eq_iEndemico}). 
It means that, 
in the case of stable DFE, 
there is no endemic fixed point, 
recovering the information previously discussed.

{As in the DFE case, 
	we analyze the stability of the EE fixed point through eigenvalues of the Jacobian matrix $\mathbb{J}_{\rm EE}$ computed on it, 
	being 
\begin{equation}
	\mathbb{J}_{\rm EE} = -\left[
	\small{
	\begin{array}{ccc}
		(\beta_0 \widetilde{i} + \lambda v + \delta + b)  & \delta  & (\beta_0 \widetilde{s} + \delta)  \\[4pt]
		-\beta_0 \widetilde{i} & (\alpha + b)  & -\beta_0 \widetilde{s} \\[4pt]
		0 & -\alpha  & (\gamma + b)
	\end{array}}
	\right].  
\end{equation} 
The correspondent characteristic polynomial is given by} 
{\begin{equation}
	\mathcal{P}(\zeta) = \zeta^3 + c_2\zeta^2 + c_1\zeta + c_0,
\end{equation}
where the coefficients are
\begin{align}
	c_0 &= (\beta_0 \widetilde{i} + \lambda v + \delta + b)\left[(\alpha + b)(\gamma + b) - \alpha\beta_0 \widetilde{s}\right] +\nonumber\\
		&+ \beta_0 \widetilde{i} \left[\alpha \left(\beta_0 \widetilde{s} + \delta \right) + \delta(\gamma + b) \right],\\
	c_1 &= (\beta_0 \widetilde{i} + \lambda v + \delta + b)(\alpha + \gamma + 2b) +\nonumber\\
		&+ (\alpha + b)(\gamma + b) + \beta_0(\delta \widetilde{i} - \alpha\widetilde{s}),\\
	c_2 &= \beta_0 \widetilde{i} + \lambda v + \alpha + \gamma + \delta + 3b.
\end{align}}
	
{Having such coefficients and using the Routh–Hurwitz criterion~\cite{Meinsma1995237}, 
	it is possible to find a parametric relationship for the EE point stability.  
	In this way, 
	the real part of the all three eigenvalues is less than zero if 
	and only if}
{\begin{equation}
		c_0, c_1, c_2 > 0;~~\text{and}~~c_0 < c_1 c_2.
\end{equation}
Fulfilled the above conditions, 
once the EE point exists, 
it is attractive.} 
{In the context of the present work, 
	we verify that the EE point is asymptotically stable for the parameters used in our simulations, 
	as already described in the previous subsection: 
	$\lambda = 1$, 
	$p=0.25$, 
	$b=0.02$, 
	$\delta=0.25$, 
	$\alpha=100$, 
	$\gamma = 100$, 
	$\beta_0 = 270$ and  $v\in[0,v_0=0.445)$, 
	where $\mathcal{R}_0>1$.}

For the non-autonomous system, 
where $\beta=\beta(t)$ is the periodic function given in Eq.~(\ref{eq_betaFuncao}), 
the DFE solution is the only one of the fixed-point type. 
\begin{figure}[!b]
	\centering
	\includegraphics[width=1.\columnwidth]{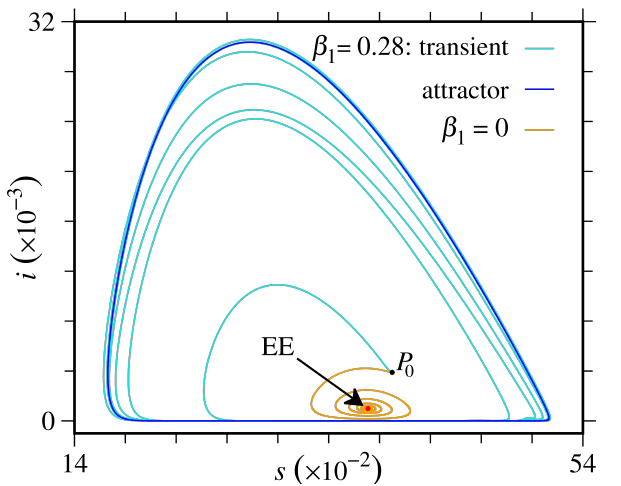}
	\caption{{Trajectories evolving from the initial condition $P_0(s_0,e_0,i_0) = (0.39,0.0039,0.0039)$ (black dot).} 
		{We consider the parameters:
			$v=0.17$,
			$\lambda = 1$, 
			$p=0.25$, 
			$b=0.02$, 
			$\beta_0 = 270$, 
			$\delta = 0.25$, 
			$\alpha = 100$ and   
			$\gamma = 100$.}
		{For $\beta_1=0$, 
			the trajectory (orange line) 
			spirals towards the attractive fixed point ${\rm EE}(0.371,0.001,0.001)$ (red dot).}
		{In the non-autonomous case ($\beta_1=0.28$ and $\omega=2\pi$), 
			the system evolves (light-blue line) around the point ${\rm EE}$ converging to the periodic attractor (blue line).}}
	\label{fig2}
\end{figure}
{In other words, 
	the fixed point solution with $i\neq0$ does not exist when the transmissivity is an explicit function of time.} 
In this case, 
endemic solutions {are not fixed points, 
but} consist of orbits that can be periodic or chaotic. 
{Figure~\ref{fig2} shows the system evolution from the initial condition $P_0(s_0,e_0,i_0) = (0.39,0.0039,0.0039)$ in the two cases:  
	1) autonomous (with $\beta_1=0$), 
	where the trajectory (orange line) spirals to the attractive fixed point 
	${\rm EE}\left(\widetilde{s},\widetilde{e},\widetilde{i}\right) \approx (0.371,0.001,0.001)$; 
	2) non-autonomous system (with $\beta_1=0.28$), 
	where the trajectory (light-blue line) evolves during a transient time around EE point and converges to the periodic attractor (blue line).} 
{To obtain these curves, 
	we adopt the aforementioned parametric configuration and 
	consider a constant vaccination rate $v=0.17$.}
In the following sections, 
we numerically investigate the impacts of different vaccination protocols on the 
dynamics of the Eqs.~(\ref{eq_sistemaNormalizado}) with seasonality, 
focusing on the infected population.

%%%%%%%%%%%%%%%%%%%%%%%%%%%%%%%%%%%%%%%%%%
%%%%%%%%%%%%%%%%%%%%%%%%%%%%%%%%%%%%%%%%%%
\section{Constant vaccination strategy} \label{sec_vacinaConstante}

\begin{figure}[!b]
	\centering
	\includegraphics[width=1.\columnwidth]{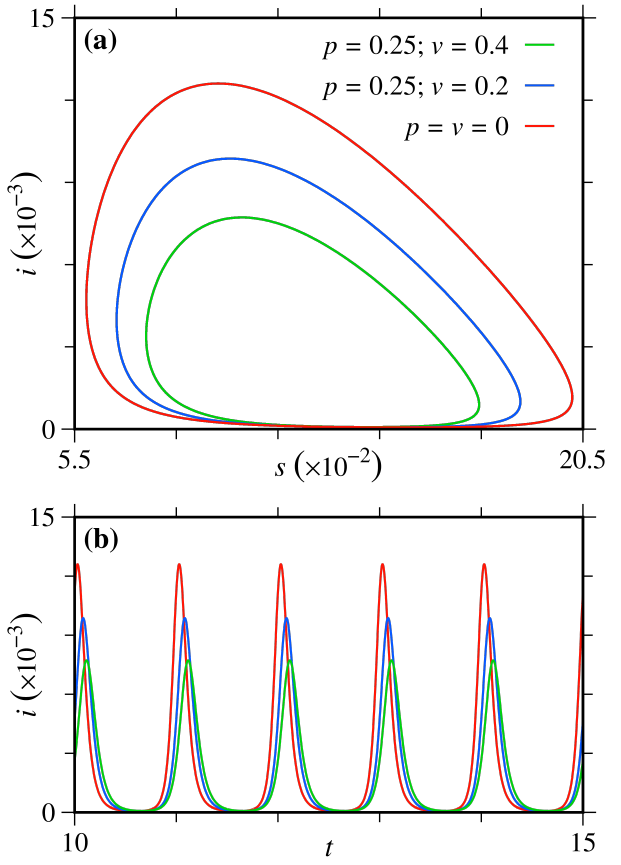}
	\caption{Numerical solutions of Eqs.~(\ref{eq_sistemaNormalizado}) for three different values of $p$ and $v$. 
		(a) Limit cycles obtained for $p=v=0$ (red curve) and 
		$p=0.25$, with $v=0.2$ (blue curve) and 
		$v=0.4$ (green curve).
		(b) Time series for $i$ corresponding to the limit cycles 
		in the same colours according to the legend.
		We consider $b=0.02$,  
		$\beta_0 = 800$, 
		$\beta_1 = 0.20$,
		$\delta = 0.25$, 
		$\alpha = 40$, 
		$\gamma = 100$, 
		$\omega=2\pi$, and
		initial condition $(s_0,e_0,i_0)=(0.9,0,0.1)$. 
		We discard $10^5$ integration steps as transient.} 
	\label{fig3}
\end{figure}
Continuous vaccination campaign is modelled by a constant vaccine term 
$v$ in Eqs. (\ref{eq_sistemaNormalizado}), {which represents 
a fraction of the host population being vaccinated at every time step. 
This approach can be used for modelled mass vaccination programs \cite{Grabenstein2006}. 
In this campaign, a certain amount of the population is vaccinated in a 
short period of time.
Some examples, are the mass campaign against measles \cite{McLean1988}, 
smallpox \cite{Fenner1988}, and COVID-19 \cite{Bagcchi2021}}.
In this case, 
$p$ is also constant. 
To numerically integrate the system of differential equations, 
we use the fourth-order Runge-Kutta method~\cite{Butcher1987} with a fixed step of $10^{-3}$. 
It is important to mention that, 
without spoiling to the analyses, 
we consider the vaccine effectiveness $\lambda=1$. 
Even so, 
this term can be understood, 
more generally, 
as absorbed by $v$, 
since these always appear as a product, 
maintaining the equations form.

Figure~\ref{fig3} displays three different numerical solutions for 
Eqs.~(\ref{eq_sistemaNormalizado}), for different $p$ and $v$. 
In these results, we consider 
$b=0.02$,  
$\beta_0 = 800$, 
$\beta_1 = 0.20$,
$\delta = 0.25$, 
$\alpha = 40$, 
$\gamma = 100$ and $\omega=2\pi$. 
{In this work, the time unity is years.}
In order to interpret these values, 
we consider $b$ as an annual birth rate, 
with the other four rates being annual too. 
Thus, 
$\omega=2\pi$ means seasonality with a period of one year. 
Regarding the vaccination parameters, 
the reference curve (red line) is obtained for $p=v=0$, 
for the other two $p=0.25$ and $v=0.2$ 
in the blue curve, 
and $v=0.4$ in the green one. 
We adopt the initial condition $(s_0,e_0,i_0)=(0.9,0,0.1)$ and 
discard the first $10^5$ integration steps as transient. 
For these configurations, 
the solutions of the system are limit cycles, 
which are projected onto the plane $i\times s$ in Fig.~\ref{fig3}(a). 
These cycles contract as the vaccination rate increase, 
with a significant decrease in the local maxima of curve $i$, 
as shown in Fig.~\ref{fig3}(b). 
We find that the presence of a vaccine campaign causes a reduction of the maximum number of infects at a time and 
this effect is amplified with the $v$ increment. 
Compared to the reference curve, 
a constant vaccination rate of one-quarter of the newborn population and 
one-fifth of the susceptible ones, 
for year, 
leads to approximately a $21.4\%$ reduction in peak infections.
Intensifying vaccination to a rate of $40\%$ of the susceptible population per year, 
this reduction reaches $\approx38.9\%$. 

For the next numerical simulations, 
based on the Gabrick's work \cite{Gabrick2023}, 
we consider the parameters 
$b=0.02$, 
$\beta_0 = 270$, 
$\beta_1 = 0.28$, 
$\delta = 0.25$, 
$\alpha = 100$, 
$\gamma = 100$ and $\omega=2\pi$. 
To compute the impact of the vaccine on the system, 
over a given finite time interval, 
we consider the ratio 
\begin{equation}
	\theta = \frac{A_{\rm v}}{A_{0}},
	\label{eq_theta}
\end{equation}
where $A_{\rm v}$ and $A_0$ are the areas under the $i$ curve with and 
without vaccine, 
respectively. Being 
\begin{align}
	A_{\rm v} &= \int\limits_{t_0}^{t_0 + \Delta t} i(t) dt,\\
	A_{0} &= \int\limits_{t_0}^{t_0 + \Delta t} i(t;p=v=0) dt,
	\label{eq_areaIntegral}
\end{align}
where
$A_0$ is the reference value. 
From this relation we conclude that
if $\theta<1$, 
then the vaccination campaign reduces the total number of infected individuals, and  
if $\theta>1$, 
the campaign has the undesired effect of increasing it. 
$\theta=1$ means that the vaccination has no effect on the total infections over time. 

\begin{figure}[!t]
	\centering
	\includegraphics[width=1.\columnwidth]{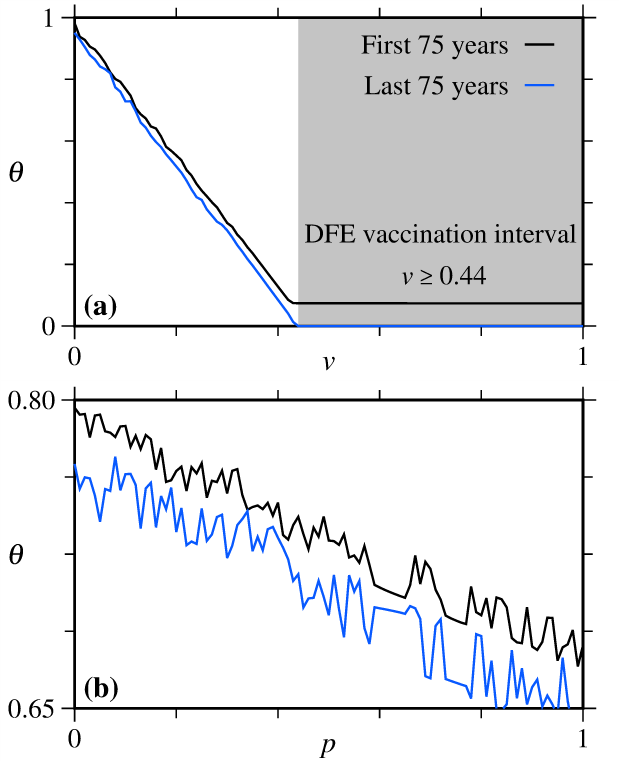}
	\caption{
		(a) $\theta$ as function of rate $v$ with $p=0.25$. 
		DFE is reached for $v=0.44$ (grey background) 
		(b) $\theta$ as function of newborn vaccine $p$ with $v=0.1$. 
		Vaccination of newborns reduced $\approx 12.5\%$ the total of infected individuals.
		Horizontal axis discredited in steps of $10^{-2}$. 
		Results of the first $75$ years without transient discard (black curve) and  
		evaluated the last $75$ years (blue curve) after $100$ years of the system evolution. 
		We consider 
		$b=0.02$, 
		$\beta_0 = 270$, 
		$\beta_1 = 0.28$, 
		$\delta = 0.25$, 
		$\alpha = 100$, 
		$\gamma = 100$ and $\omega=2\pi$; 
		initial condition $(s_0,e_0,i_0)=(0.9,0,0.1)$. } 
	\label{fig4}
\end{figure}

Figure~\ref{fig4} shows the influence of vaccination in two different strategies. 
In panel (a) we fixed the fraction of newborns $p=0.25$ and 
get $\theta$ as a function of the vaccination rate $v\in[0,1]$, 
with the horizontal axis discredited in steps of $\Delta v = 10^{-2}$. 
We evolve the system from the same initial condition used in Fig.~\ref{fig3} and 
calculate $\theta$ both for the first $75$ years without transient discard (black curve) and, 
past $100$ years of evolution, 
for the last $75$ years (blue curve). 
Without discarding the early years, 
$\theta$ presents a linear decay as a function of $v$, increasing until $v\approx 0.44$, 
and stabilising in $\theta\approx0.075$. 
This value is due to the nonzero area associated with the spread of infection from the initial condition. 
In the blue curve we see that the DFE is reached for $v\geq0.44$ (grey background). 
Furthermore, it is worth mention that in our simulation if we consider a 
large number of years, e.g., 100, 125 years, the DFE also is reached for $v\geq0.44$. 
Figure~\ref{fig4}(b) displays $\theta$ as a function of $p$, with $v=0.1$.
This result shows when whole newborns are vaccinated occurs a reduction 
$\approx12.5\%$ of the total infected individuals.  

\begin{figure}[!t]
	\centering
	\includegraphics[width=1.\columnwidth]{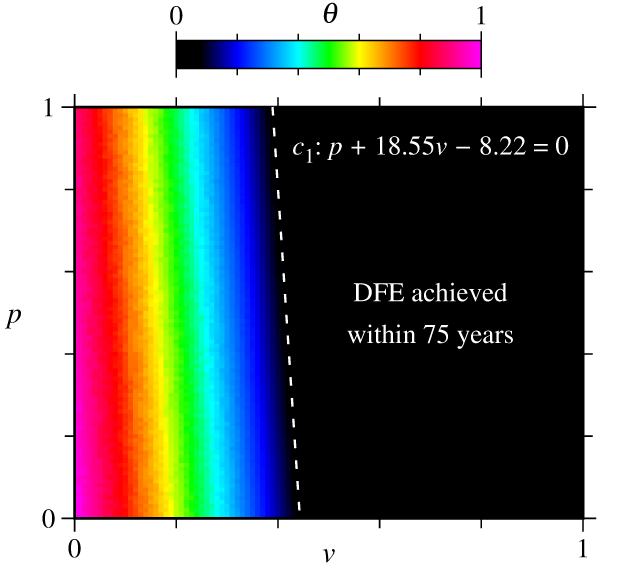}
	\caption{Parameter plane $p \times v$ with {$\theta$} in colour scale. 
		Both axis discretised in steps of $10^{-2}$. 
		We consider 
		$b=0.02$, 
		$\beta_0 = 270$, 
		$\beta_1 = 0.28$, 
		$\delta = 0.25$, 
		$\alpha = 100$, 
		$\gamma = 100$ and 
		$\omega=2\pi$.
		$\theta$ is calculated for the first 75 years of infection, 
		without transient discard and with initial condition $(s_0,e_0,i_0)=(0.9,0,0.1)$.
		DFE occurs from the dashed white line ($c_1$) to the right of it, 
		where $p \geq 8.22 - 18.55v$.} 
	\label{fig5}
\end{figure}

We evaluate the effects of combining different newborns and 
susceptible vaccination rates in the 
parameter plane $p \times v$, 
with $\theta$ in colour scale, 
displayed in Fig.~\ref{fig5}. 
Our results show the prevalence of $v$ over $p$ in reducing the spread of infection. 
This is due to the intrinsic characteristics of the model, 
if $\delta$ decreases, 
the parameter $p$ becomes more relevant. 
DFE is achieved within $75$ years from the onset of infection by $(v,p)$ pairs from 
the dashed white line to the right of it, 
defined by the straight line equation $c_1: p + 18.55v - 8.22 = 0$. 
Extinction of infection in the host population occurs when $e = i = 0$ and 
can be verified in simulations with a given numerical precision. 
In order to determine the boundary line $c_1$ more accurately, 
we integrate the system in the region defined by $v\in[0.39,0.46]$ and 
$p\in[0,1]$ in a uniform grid of $201\times201$ points. 
Still in Fig.~\ref{fig5}, 
the blue band indicates that the constant vaccination strategy leads 
to a significant reduction of the total infections, 
computed around $20\%$ of the reference. 
In the green band, 
where $v\approx 0.2$, 
the reduction is around $50\%$. 
For small vaccination rates, 
with $v\approx10\%$ and less, 
the amount of infections decreases only slightly, 
being around $70\%$ to $80\%$ (gradient from orange to red) of the result without vaccination, 
even larger than $90\%$ (magenta band). 

\begin{figure}[!b]
	\centering
	\includegraphics[width=1.\columnwidth]{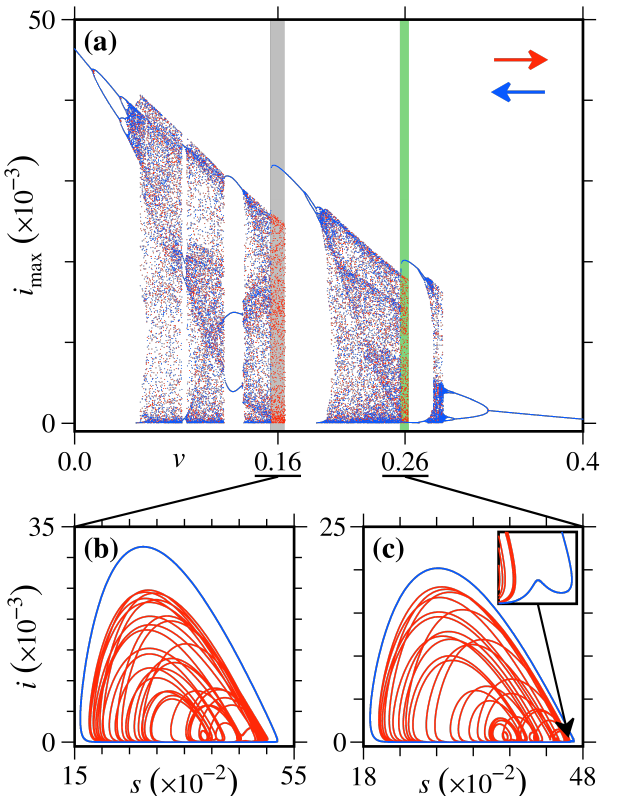}
	\caption{(a) Hysteresis type bifurcations diagram (HTBD) for $i$ local maxima values ($i_{\rm max}$) as 
		function of $v\in[0,0.4]$ discretised in steps of $4\times 10^{-4}$. 
		We consider 
		$b = 0.02$, 
		$\beta_0 = 270$, 
		$\beta_1 = 0.28$, 
		$\delta = 0.25$, 
		$\alpha = 100$, 
		$\gamma = 100$, 
		$\omega = 2\pi$ and $p = 0.25$. 
		For each value of $v$, 
		a transient of $10^5$ integration steps is discarded and 
		computed $i_{\rm max}$ over the last $75$ years of the infection. 
		The red points are in the forward $v$ direction and blue points in the backward direction.
		Grey and green backgrounds highlight bi-stability intervals. 
		{Projections in the plane $i\times s$ of periodic (blue line) and 
			chaotic (red line) attractors coexisting for 
			(b) $v = 0.16$ and 
			(c) $v = 0.26$.}} 
	\label{fig6}
\end{figure}

The vaccination does not only affects the number of infected individuals, 
but also the dynamics. 
Such effects are examined by hysteresis type bifurcations diagrams (HTBD), 
as shown in Fig. \ref{fig6}. 
This kind of bifurcation diagram is generated by evolving the system 
in a given discretized interval of a control parameter in both directions along the horizontal axis, 
first in its growth (red points), 
then in decrease (blue points), 
assuming the final state of the system at the current parameter value as the initial condition for the next one. 
This numerical technique is especially useful for finding bi-stability, 
when it exists \cite{Medeiros2017}. 
Figure \ref{fig6}(a) displays a HTBD of $i$ local maxima values ($i_{\rm max}$) 
as a function of the susceptible vaccination rate.
To compute the peaks in the $i$ time series, 
we discard the first $10^5$ integration steps as transient and 
evaluate the evolution of the infection over the last $75$ years in the simulation. 
For the considered parameters, the bi-stability dynamics 
between chaotic and periodic attractors exist without vaccination \cite{Gabrick2023}. 
The inclusion of vaccination is able to destroy the bi-stability for some 
parameters values. 
However, 
the bi-stability remains in two ranges of the control parameter, 
highlighted in Fig.~\ref{fig6}(a), 
by the grey and 
green background columns. 
In the approximate interval $v\in(0.155,0.165)$ (grey background) 
the periodic orbit (blue dots) 
has a maximum value $i_{\rm max}\approx0.032$, 
{while} the chaotic one (red dots) 
has a smaller maximum of $i_{\rm max}\approx0.026$. 
Also in the interval $v\in(0.256,0.263)$ (green background), 
we observe the periodic orbit with an extreme value of $i_{\rm max}$ 
slightly higher than that seen from the chaotic one. 
{Chaotic behaviour can be related to non-predictability in infectious diseases spread~\cite{Gabrick2023}. 
Prior knowledge of these behaviors or how they can be modified is crucial for epidemic control strategies.} 
{Figure~\ref{fig6}(b) and \ref{fig6}(c) display the $i \times s$ 
	projection of the attractors obtained for the highlighted values in the bi-stability for, 
	respectively, $v = 0.16$ and 
	$v = 0.26$.}
{Both plots show the periodic solution (blue line) 
	with a higher peak than the chaotic one (red line).} 
{For $v = 0.16$, 
	we observe a chaotic attractor from the initial condition $P_1(s_0,e_0,i_0) = (0.9,0,0.1)$, 
	while the periodic one emerges for $P_2(s_0,e_0,i_0) = (0.3,0.03,0.03)$.}
{However, 
	for $v = 0.26$, 
	$P_1$ leads to the periodic attractor and  
	$P_2$ to the chaotic one.} 
{The magnification in the panel (c) exhibits a small amplitude local maximum of $i$, 
	occurring before the large increase in infectious cases and 
	showing the two maxima as obtained in the bifurcation diagram.}

%%%%%%%%%%%%%%%%%%%%%%%%%%%%%%%%%%%%%%%%%%
%%%%%%%%%%%%%%%%%%%%%%%%%%%%%%%%%%%%%%%%%%
\section{Pulsed vaccination strategy} \label{sec_vacinaPulsada}
{Periodic vaccination consists of the appliance of 
vaccine in a certain fraction of the population in periodic time intervals.
Examples of this strategy is considered in seasonal 
influenza \cite{Alkhamis2022} and in control of childhood viral infections, 
such as measles and polio \cite{Nokes1995}.}
\begin{figure}[!b]
	\centering
	\includegraphics[width=1.\columnwidth]{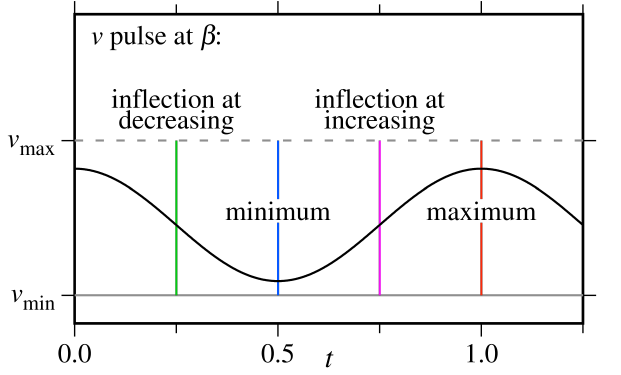}
	\caption{Schematic illustration of pulsed $v$ with 
		maximum $v_{\rm max}$ (dashed grey line level), 
		and minimum level $v_{\rm min}$ (solid grey line). 
		Pulses in four different timings relative to $\beta(t)$ (black curve): 
		at inflections (green and magenta vertical lines) and 
		at the local minimum and maximum, blue and red lines, respectively.} 
	\label{fig7}
\end{figure}
In our modelling approach, we consider
this immunisation strategy consists of maintaining a minimum level $v=v_{\rm min}$ 
during the year and raising the vaccination rate to $v=v_{\rm max}\geq v_{\rm min}$ in 
concentrated pulses at specific times according to the oscillation of the transmissivity. 
An illustrative scheme is shown in Fig. \ref{fig7}, 
where different pulse timings are superimposed on the periodic curve $\beta(t)$ (black line). 
We consider four variants of the same strategy, 
in which the pulse occurs,
relative to the $\beta(t)$, 
at every time corresponding to the: 
inflection point at decreasing (green line); 
local minimum (blue line);
inflection point at the curve growth (magenta line); and 
local maximum (red line). 
Being the vaccination rate  
\begin{align}
	v(t;\tau) = 
	\begin{cases}
		v_{\rm max},~\text{if}~~\{t - \tau\} = 0,\\
		v_{\rm min},~\text{otherwise}, 
	\end{cases}
	\label{eq_vPulso}
\end{align}
where $\{t - \tau\}$ is the non-integer part of $(t - \tau)$. 
For the first variant of the strategy we have $\tau = 0.25$, 
in the second $\tau = 0.5$, 
next $\tau = 0.75$ and in the last one $\tau = 0$.
 
Considering a time interval of $\approx 9$ hours in one day per year, 
we investigate the impact on total infected in the first $75$ years of infection, 
with $v_{\rm min}=0$ and $v_{\rm max}\in[0,1000]$, 
where $v=1000$ means that the entire susceptible population is vaccinated in one pulse,  
since the integration step is $10^{-3}$ and 
the proportion of people vaccinated is given by the product of the campaign duration with the vaccination rate. 
Furthermore, 
we adopt $p=0$ and 
the same initial condition used in the previous section.

\begin{figure}[!b]
	\centering
	\includegraphics[width=1.\columnwidth]{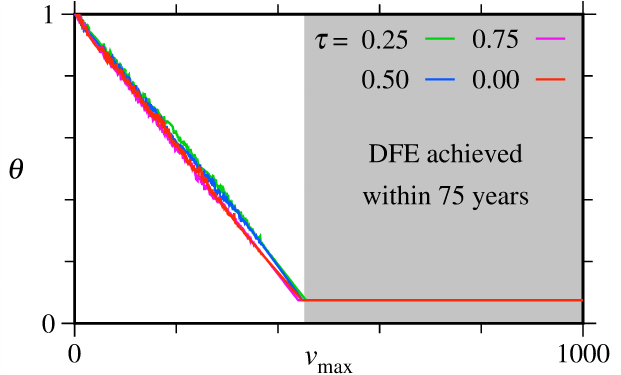}
	\caption{$\theta$ as a function of $v_{\rm max}\in[0,1000]$ 
		from four variants of the pulsed vaccination strategy and 
		obtained in the first $75$ years of infection. 
		Horizontal axis discretised in $1001$ equidistant points. 
		We consider 
		$b = 0.02$, 
		$\beta_0 = 270$, 
		$\beta_1 = 0.28$, 
		$\delta = 0.25$, 
		$\alpha = 100$, 
		$\gamma = 100$, 
		$\omega = 2\pi$ and	$p = v_{\rm min} = 0$. 
		Initial condition $(s_0,e_0,i_0)=(0.9,0,0.1)$. 
		Pulse application timing identified in colours, 
		according to the legend. 
		DFE is achieved (grey background) within $75$ years for $v_{\rm max} \geq 452$ with $\tau \in \{0, 0.5\}$, 
		$v_{\rm max} \geq 459$ for $\tau = 0.25$ and 
		$v_{\rm max} \geq 444$ when $\tau = 0.75$.} 
	\label{fig8}
\end{figure}
Figure~\ref{fig8} displays $\theta$ as a function of $v_{\rm max}$, 
with the horizontal axis discretised in steps of $\Delta v_{\rm max}=1$, 
equivalent to increments of $0.1\%$ in the reach of the immunisation campaign. 
The four variants of the strategy are covered, 
with the pulse applied at the first (green) and 
second (magenta) inflections of $\beta(t)$ and, 
complementary, 
for both $\tau=0.5$ (blue) and $\tau=0$ (red), 
at times corresponding of the transmissivity minimum and 
maximum, respectively. 
These results are practically the same. 
There is no expressive difference as to when a concentrated immunisation campaign works, 
with a slight advantage in applying the pulse at the local maximum ($\tau = 0$) or 
second inflection ($\tau = 0.75$) of the transmissivity curve. 
When the campaign coincides with the extremes of the $\beta(t)$ curve ($\tau=0$ and $\tau=0.5$), 
DFE is achieved within the initial $75$ years (grey background) 
from $v_{\rm max} = 452$, 
i.e. 
vaccinating from $45.2\%$ of the susceptible population on a single day per year. 
This threshold is $49.0\%$ for $\tau=0.25$ and $44.4\%$ with $\tau=0.75$, 
the latter being the best result. 
\begin{figure}[!b]
	\centering
	\includegraphics[width=1.\columnwidth]{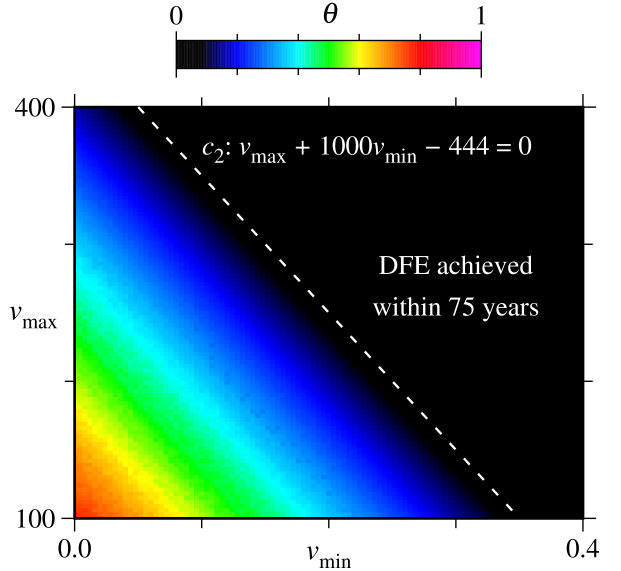}
	\caption{Parameter plane $v_{\rm max}\times v_{\rm min}$ discretised on a uniform grid of $101\times101$ points, 
		with $\theta$ (colour scale) calculated for the first $75$ years of infection. 
		We consider 
		$b = 0.02$, 
		$\beta_0 = 270$, 
		$\beta_1 = 0.28$, 
		$\delta = 0.25$, 
		$\alpha = 100$, 
		$\gamma = 100$, 
		$\omega = 2\pi$, 
		$p = 0$ and $\tau=0.75$. 
		Initial condition $(s_0,e_0,i_0)=(0.9,0,0.1)$.
		DFE occurs from the dashed white line ($c_2$) to the right of it, 
		where $v_{\rm max} \geq 444 - 1000v_{\rm min}$.}
	\label{fig9}
\end{figure}

According to our simulations, 
considering a not null baseline $v_{\rm min}>0$, 
it is possible to reduce the pulse amplitude and 
yet result in the extinction of the disease in the host population within $75$ years. 
In Fig. \ref{fig9}, we show the parameter plane $v_{\rm max}\times v_{\rm min}$ reinforcing this proposal. 
We define $\tau = 0.75$, 
applying the pulse at the moment of inflection in increasing of the transmissivity, 
varying $v_{\rm min}\in[0,0.4]$ and 
$v_{\rm max}\in[100,400]$ in a uniform grid of $101\times101$ points. 
We highlight the white dashed line $c_2: v_{\rm max} + 1000v_{\rm min} - 444 =0$, 
from which we obtain DFE. 
Pairs $(v_{\rm min},v_{\rm min})$ in the blue band bring the total infected to 
$\approx20\%$ of those that would occur without vaccination, 
e.g., 
how we get with the base rate $v_{\rm min}\approx 0.28$ and 
the intensive campaign $v_{\rm max}=100$, 
the latter meaning vaccination of $10\%$ of the susceptible population. 
Since the base value throughout the year is around $v_{\rm min}=0.125$, 
even with $v_{\rm max}=100$, 
the number of people infected during the $75$ years simulated is reduced by approximately half (green band), 
when compared to the reference case. 

Along lines parallel to $c_2$, 
the proportion of susceptible vaccinated annually is a constant given by
\begin{equation}
	\rho_{\rm line} = 0.001v_{\rm max} + v_{\rm min}.
\end{equation} 
Thus, 
on line $c_2$ we have $\rho_{c_2}=0.444$, 
close to the DFE threshold value in line $c_1$ with $p=0$, 
shown in Sec. \ref{sec_vacinaConstante}. 
In these simulations, 
an immunisation campaign that reaches $\approx 22.5\%$ of susceptible annually reduces to $50\%$ the accumulated of infected people, 
whereas a rate of $\approx 38\%$ causes a decrease of $80\%$. 
Similar values to those obtained with $v$ constant and without newborns vaccination, 
displayed in Fig. \ref{fig5}. 
The strategy of intensifying immunisation in pulses allows the reduction at baseline, 
being $v_{\rm min}$ less than the rates required for the same results with constant vaccination rate. 

%%%%%%%%%%%%%%%%%%%%%%%%%%%%%%%%%%%%%%%%%%
%%%%%%%%%%%%%%%%%%%%%%%%%%%%%%%%%%%%%%%%%%
\section{Pulsed width vaccination strategy} \label{sec_vacinaPulsoLargo}
Extending the proposal of the previous section, 
we modify the pulsed vaccination strategy considering, 
now, 
a longer duration of the intensified campaign.
Similarly to the formulation of Eq. (\ref{eq_vPulso}), 
the pulse is centred at $t = \tau + k$, 
$\forall~k\in\mathbb{Z}$, 
but with a width of $D\in[0,1]$, 
which corresponds to the duty cycle. 
Outside the intensive campaign interval, 
vaccination assumes a baseline rate $v_{\rm min}$, 
according to the following mathematical description
\begin{equation}
	v(t;\tau) = 
	\begin{cases}
		v_{\rm max},~\text{if}~~\left\{(t - \tau) + \dfrac{D}{2} \right\} \leq D,\\
		v_{\rm min},~\text{otherwise}.
	\end{cases}
	\label{eq_vPulsoLargo}
\end{equation}
\begin{figure}[!b]
	\centering
	\includegraphics[width=1.\columnwidth]{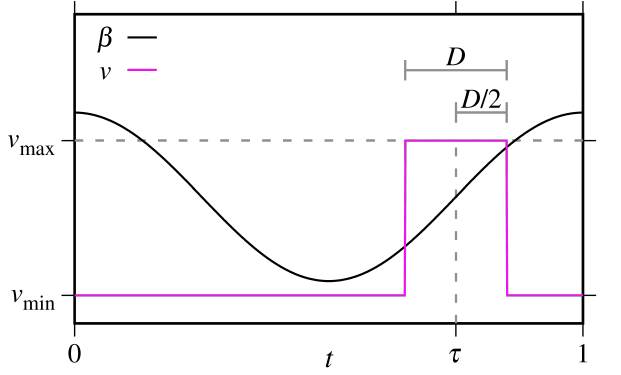}
	\caption{Schematic illustration of pulsed width vaccination. 
		Transmissivity $\beta(t)$ (black curve) superimposed by vaccination rate $v(t)$ (magenta curve), 
		it is on two levels:		
		maximum $v_{\rm max}$ (dashed grey line level) and minimum $v_{\rm min}$. 
		Pulse width $D$ centred on $\tau$, 
		which is a time instant relative to the seasonal cycle.}
	\label{fig10}
\end{figure} 
Remembering that $\{x\} \in [0,1)$ is the non-integer part of the real number $x$. 
Figure~\ref{fig10} displays a schematic illustration of this strategy, 
where the seasonal transmissivity (black curve) is superimposed on the 
time-dependent vaccination rate curve (magenta curve). 
For instance, 
in the representation $D = 0.2$, 
corresponds to one fifth of the year. 
The concentrated pulse strategy, 
according Eq. (\ref{eq_vPulso}), 
is recovered when $D=0$. 
As for $D=1$ we return to the constant vaccination rate. 
Based on the best result presented in Sec. \ref{sec_vacinaPulsada}, 
in the next simulations we adopt $\tau=0.75$.

\begin{figure}[!b]
	\centering
	\includegraphics[width=1.\columnwidth]{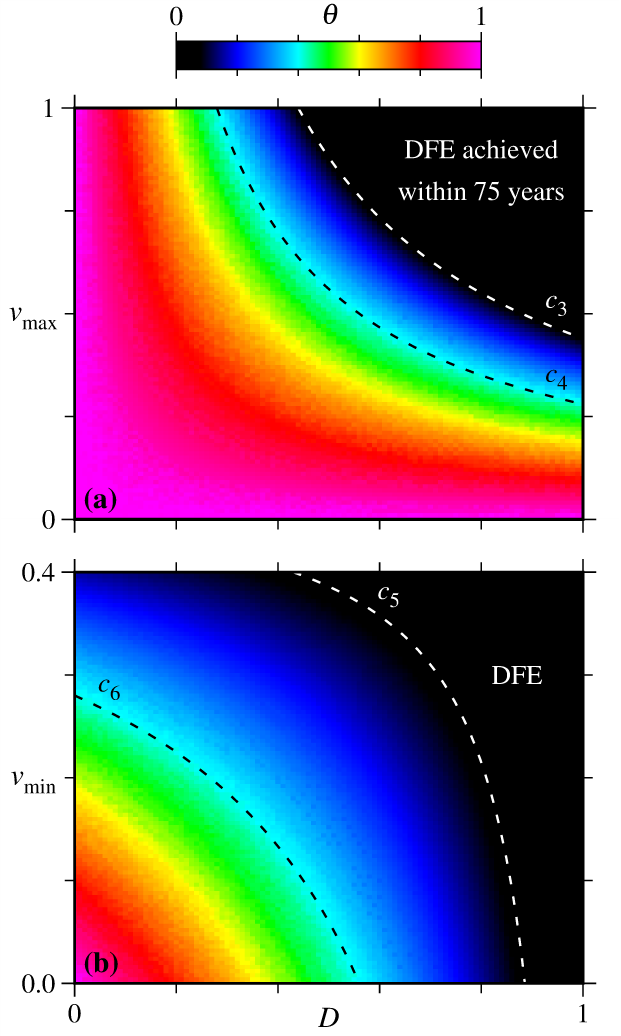}
	\caption{Parameter planes discretised on a uniform grid of $101\times101$ points, 
		with $\theta$ (colour scale) calculated for the first $75$ years of infection. 
		(a) Plane $v_{\rm max}\times D$ with $v_{\rm min} = 0$ fixed. 
		DFE threshold $c_3:Dv_{\rm max}=0.44$ and 
		$\theta\approx0.4$ along the curve $c_4:Dv_{\rm max}=0.28$ (black dashed line). 
		(b) Plane $v_{\rm min}\times D$ with $v_{\rm max} = 0.5$ fixed.
		DFE threshold $c_5:v_{\rm min}=(0.44 - 0.5D)/(1 - D)$ and 
		$\theta\approx0.4$ along the curve $c_6:v_{\rm min}=(0.28 - 0.5D)/(1 - D)$ (black dashed line).
			We consider 
		$b = 0.02$, 
		$\beta_0 = 270$, 
		$\beta_1 = 0.28$, 
		$\delta = 0.25$, 
		$\alpha = 100$, 
		$\gamma = 100$, 
		$\omega = 2\pi$, 
		$p = 0$ and $\tau=0.75$. 
		Initial condition $(s_0,e_0,i_0)=(0.9,0,0.1)$. 
		In both panels, 
		DFE occurs from the dashed white lines ($c_3$ and $c_5$) to the right, 
		where $\rho \geq 0.44$.} 
	\label{fig11}
\end{figure}

First, 
we keep $p=0$ and 
obtain $\theta$ in the parameter planes shown in Fig. \ref{fig11}. 
As in the previous sections, 
we take into account the first $75$ years of infection and 
consider the same initial condition. 
Both panels display uniform grids of $101\times101$ points. 

We analyse the results taking into account the annual reaching of the immunisation campaign, 
given by 
\begin{equation}
	\rho = v_{\rm min} + D (v_{\rm max} - v_{\rm min}), 
	\label{eq_rhoAnual}
\end{equation}
	and the proportion of vaccinated susceptible during each pulse, 
	that is 
\begin{equation}
	\rho_{\rm pulse} = D v_{\rm max}. 
	\label{eq_rhoPulso}
\end{equation}

In Fig. \ref{fig11}(a) we fix the baseline $v_{\rm min} = 0$ and 
vary the two remaining parameters of Eq.~(\ref{eq_vPulsoLargo}) in the range $0\leq D,v_{\rm max}\leq 1$. 
In this configuration $\rho=\rho_{\rm pulse}$ and 
the DFE threshold is obtained from $\rho_{\rm pulse}=0.44$, 
i.e. 
from the curve $c_3:Dv_{\rm max}=0.44$ to the right of it. 
Already $\rho_{\rm pulse} = 0.28$ leads to $\theta\approx0.4$ (cyan band), 
value obtained along the curve $c_4: Dv_{\rm max}=0.28$ (dashed black line). 
Similar relationships can be obtained for the other $\theta$ values, 
with the balance between pulse height and 
duty cycle being the determining characteristic for reducing the number of infected people over the $75$ years simulated. 
Figure \ref{fig11}(b) illustrates the parameter plane $v_{\rm min} \times D$ with $v_{\rm max} = 0.5$, 
where we evaluate the baseline vaccination interval $0\leq v_{\rm min}\leq 0.4$. 
As evidenced in panel (a), 
the constant $\theta$ curves are determined by the proportion of susceptible individuals vaccinated during a year, 
so we get them from Eq. (\ref{eq_rhoAnual}). 
We check that DFE is reached from the curve $c_5:(0.44 - 0.5D)/(1 - D)$ to the right, 
where $44\%$ of susceptible are vaccinated annually. 
The total infected are reduced to $\approx 40\%$ around the curve $c_6:(0.28 - 0.5D)/(1 - D)$ (black dashed line), 
being $28\%$ of susceptible vaccinated along each year. 

Inclusion of newborns immunisation changes the minimum proportion of susceptible immunised annually that leads to DFE.
In order to obtain the boundary surface $c_{\rm DFE}$ as a function of $p$,
we use the equation of the line $c_1$, 
described in Sec. \ref{sec_vacinaConstante}. 
Assuming a constant vaccination rate $v=\rho_{\rm DFE}$ along that line and 
relating to Eq. (\ref{eq_rhoAnual}), 
we infer 
\begin{equation}
	 v_{\rm min} + D(v_{\rm max} - v_{\rm min}) - \frac{8.22 - p}{18.55} = 0. 
\end{equation}
DFE is achieved, during the simulated $75$ years, once
\begin{equation}
	v_{\rm min} + D(v_{\rm max} - v_{\rm min}) \geq \frac{8.22 - p}{18.55}. 
\end{equation}
This relation recovers the equations obtained for the thresholds 
$c_1$, $c_2$, $c_3$ and 
$c_5$ presented throughout the text. 
Note that in the plane, 
$c_{\rm DFE}$ reduces to curves. 

\begin{figure}[!b]
	\centering
	\includegraphics[width=1.\columnwidth]{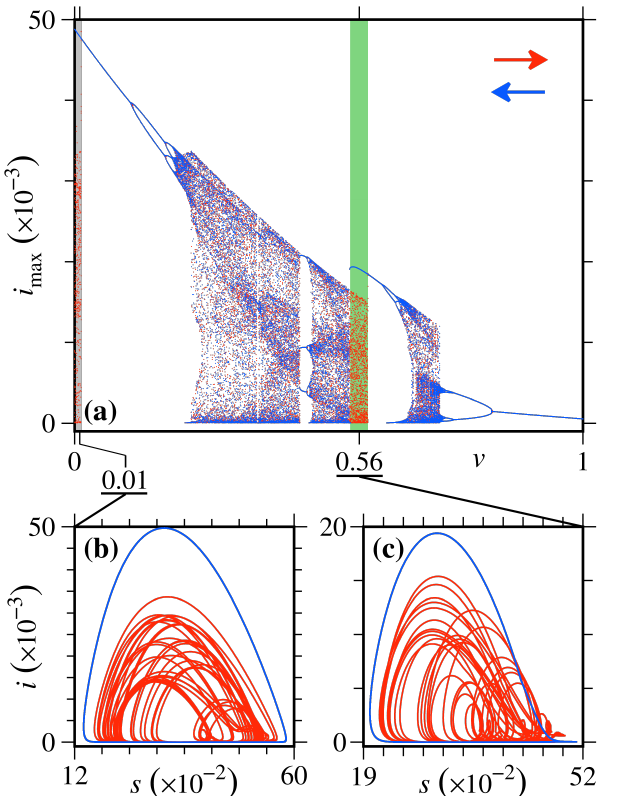}
	\caption{(a) Hysteresis type bifurcations diagram (HTBD) for $i$ local maxima values ($i_{\rm max}$) as 
		function of $v_{\rm max}\in[0,1]$ discretised in steps of $10^{-3}$. 
		We consider 
		$b = 0.02$, 
		$\beta_0 = 270$, 
		$\beta_1 = 0.28$, 
		$\delta = 0.25$, 
		$\alpha = 100$, 
		$\gamma = 100$, 
		$\omega = 2\pi$, 
		$p = v_{\rm min} =0$ and $D=0.4$. 
		For each value of $v_{\rm max}$, 
		a transient of $10^5$ integration steps is discarded and 
		computed $i_{\rm max}$ over the last $75$ years of the infection. 
		Red points are in the forward $v_{\rm max}$ direction and blue ones in the backward direction. 
		Grey and green backgrounds highlight bi-stability intervals.
		{Projections in the plane $i\times s$ of periodic (blue line) and 
			chaotic (red line) attractors coexisting for 
			(b) $v = 0.01$ and 
			(c) $v = 0.56$.}} 
	\label{fig12}
\end{figure}  

In addition, 
the bi-stability depends on the pulsed vaccination parameters. 
Figure \ref{fig12}(a) displays a HTBD of infected local maxima values ($i_{\rm max}$) 
as a function of $v_{\rm max}\in[0,1]$, 
with $p = v_{\rm min}=0$ and $D=0.4$. 
The results yielded in forward direction are red points and 
backward ones are the blue points. 
The peaks are computed discarding the first $10^5$ integration steps as transient and 
evaluate the $i$ series over the last $75$ years in the simulation. 
Similarly to the case of a constant vaccination rate, 
shown in Fig. \ref{fig6}(a), 
there are two ranges (highlighted backgrounds) 
of the control parameter where periodic and chaotic orbits coexist. 
For $v_{\rm max}\in[0,0.014)$ (grey background) 
the periodic orbit (blue dots) has a maximum value $i_{\rm max}\approx0.049$,  
However, the chaotic one (red dots) has a smaller maximum of $i_{\rm max}\approx0.036$,  
{as also showed by the attractors projection in Fig.~\ref{fig12}(b) 
	for $v_{\rm max} = 0.01$.}
In the next interval $v_{\rm max}\in(0.542,0.577)$ (green background), 
the periodic orbit presents a maximum value $i_{\rm max}\approx0.019$, 
while the chaotic one has an extreme $\approx0.015$, 
i.e., 
the periodic solution has peak infections values $26.7\%$ above the chaotic case. 
{This difference between the peak of infectious is better observed in the $i \times s$ projection of the attractors, 
	as displayed in Fig.~\ref{fig12}(c), 
	for $v_{\rm max}= 0.56$.}
{The chaotic solutions (red line) in both Fig.~\ref{fig12}(b) and \ref{fig12}(c)  
	are obtained from the initial condition $P_1$, 
	the periodic attractors (blue line) are from (b) $P_3(s_0,e_0,i_0) = (0.27,0.05,0.05)$ and 
	(c) $P_4(s_0,e_0,i_0) = (0.28,0.02,0.02)$}.

%%%%%%%%%%%%%%%%%%%%%%%%%%%%%%%%%%%%%%%%%%%%%%%
%%%%%%%%%%%%%%%%%%%%%%%%%%%%%%%%%%%%%%%%%%%%%%%
\section{Conclusions} \label{sec_conclusion}
In this work, 
we study a SEIRS model with seasonal transmissivity and vaccination control. 
Considering an autonomous version of the system, 
we obtain two equilibrium points: 
the DFE and 
the endemic fixed point. 
For the DFE, 
we established a limit for the vaccination rate as function of the other parameters in the model. 
We find that the endemic solution can only exist if the DFE is not stable.
By numerical simulations, 
we explore three different vaccination strategies in the non-autonomous case: 
constant, 
pulsed and pulsed width vaccination strategy.
All are based on immunising a proportion of susceptible individuals, 
as well as a fraction of newborns. 
In order to evaluate these three different strategies, 
we analyse the accumulated infected over a simulated interval of $75$ years:

$i$) The first one refers to a constant vaccination rate applied to the susceptible individuals. 
Our results show that the immunisation of newborns is able to slightly reduce the 
total number of infected in the simulated time interval. 
However, vaccination of susceptible is more efficient, 
with a constant rate of $\approx44\%$, 
or greater, 
leading to the extinction of the infection in the host population. 
We also identify the occurrence of bi-stability as a function of the vaccination rate, 
with periodic orbits presenting values of infected higher than the chaotic ones. 

$ii$) The second strategy is the pulsed vaccination, 
in which the immunisation campaign operates intensively on a single day per year, 
more precisely and according to the simulations, 
just $9$ hours on one day in each year. 
The baseline vaccination rate is a small constant. 
Pulses are applied annually at timings given in relation to seasonality, 
thus modelling the campaign acting according to the transmissivity variation. 
The timing of the immunisation pulse does not lead to a significant difference in the accumulated number of infected. 
Even so, 
acting at the inflection point of the rise in the transmissivity curve proved to be slightly more effective, 
reducing the percentage of susceptibles vaccinated during the pulse to result in DFE. 
The baseline and intense immunisation rate has a linear relationship with the reduction of disease spread.  
Hence, the total number of cases depends on the proportion of susceptible vaccinated annually, and 
not on independent immunisation rates. 

$iii$) In our last strategy, we applied a pulsed width vaccination campaign. 
We discover that the disease spread depends on a non-linear relationship between vaccination rates and 
the duty cycle of the campaign. 
We derive the relationship between the immunisation parameters that leads to the extinction of the infection within the simulated $75$ years. 
 It depends on the baseline and intensive vaccination rates, 
the campaign duty cycle and the proportion of immunised newborns. 
We find that the reduction of the accumulated infected number depends on the annual vaccination rate of the susceptible population, 
this finding being valid for all three strategies, 
mainly in the first one. 
For example, 
regardless of the strategy, 
given a vaccination of $\approx38\%$ of susceptible each year, 
the cumulative case of infection is reduced to only $20\%$ of what it would be in a no-vaccination situation. 
Since we study the model with normalised variables, 
we are always dealing with proportions of the total population. 
For the constant population approximation, 
a direct transition to the number of infected people over time is valid, 
whereas for the more general case, 
where the population may vary, 
we must take the results into account as proportions. 

We plan to study, 
in future works, 
the effects of periodic and 
perturbed vaccination in chaotic dynamics and 
bi-stable parameter rages. 
In addition, 
we plan to fill some points opened in this work, 
which is exploring the effects of stochastic process in vaccination campaign. 

%%%%%%%%%%%%%%%%%%%%%%%%%%%%%%%%%%%%%%%%%%%%%%%
\section*{Acknowledgements}
The authors thank the financial support from the Brazilian Federal Agencies (CNPq); 
S\~ao Paulo Research Foundation (FAPESP) under Grant Nos. 2021/12232-0, 
\newline2018/03211-6, 
2022/13761-9;
Coordenação de Aperfeiçoamento de Pessoal de Nível Superior (CAPES);  Funda\-\c c\~ao A\-rauc\'aria. 
R.L.V. received partial financial support from the  following Brazilian government agencies: CNPq (403120/2021-7, 301019/2019-3), CAPES (88881.143103/2017-01), FAPESP (2022/04251-7). 
E.C.G. received partial financial support from
Coordenação de Aperfeiçoamento de Pessoal de Nível Superior - Brasil (CAPES) - Finance Code 88881.846051/2023-01.
We thank 105 Group Science (www.105groupscience.com). 

%%%%%%%%%% References %%%%%%%%%%%%%%
\section*{References} %Gambiarra

\end{document}